\documentclass{article}

\usepackage[dvips]{graphicx}
\usepackage{amsmath}

\title{iMet: A computational tool for structural annotation of 
unknown metabolites from tandem mass spectra}

\author{
  Antoni Aguilar-Mogas\thanks{Departament d'Enginyeria Qu\'imica,
                          Universitat Rovira i Virgili,
                          Tarragona, Spain.}
  \and 
  Marta Sales-Pardo\footnotemark[1]
  \and
  Miriam Navarro\thanks{Centre for Omic Sciences, 
                                   Universitat Rovira i Virgili, 
                                   Reus, Spain.}
                 \thanks{Metabolomics Platform,
                                        Spanish Biomedical Research Center
                                        in Diabetes and Associated Metabolic
                                        Disorders (CIBERDEM),
                                        Madrid, Spain.}
  \and
  Ralf Tautenhahn\thanks{Scripps Center for Metabolomics 
                                        and Mass Spectrometry,
                                        The Scripps Research Institute,
                                        La Jolla, CA, USA.}
  \and
  Roger Guimer\`a\footnotemark[1] \thanks{Instituci\'o Catalana de Recerca 
                                  i Estudis Avan\c{c}ats (ICREA),
                                  Barcelona, Spain.}
                                  \thanks{To whom correspondence 
                                  should be addressed: roger.guimera@urv.cat; oscar.yanes@urv.cat}
  \and
  Oscar Yanes\footnotemark[2] \footnotemark[3] \footnotemark[6]
  }

\date{}

\begin{document}

\maketitle

\begin{abstract}
Untargeted metabolomic 
studies are revealing large numbers of naturally occurring metabolites that 
cannot be characterized because their chemical structures and MS/MS spectra 
are not available in databases. 
Here we present iMet, a computational tool based on experimental tandem mass 
spectrometry that allows the annotation of metabolites not discovered 
previously. iMet uses MS/MS spectra to identify metabolites structurally 
similar to an unknown metabolite, and gives a net atomic addition or removal 
that converts the known metabolite into the unknown one. We validate the 
algorithm with 148 metabolites, and show that for 89\% of them at least one 
of the top four matches identified by iMet enables the proper annotation of 
the unknown metabolite. 
iMet is freely available at http://imet.seeslab.net.
\end{abstract}

\section{Introduction}
The great success in the characterization of genes, 
transcripts and proteins is a direct consequence of two factors.
First, such molecules result from the concatenation of a small set of known 
monomers, namely, nucleotides and amino acids. Second, existing technologies 
and bioinformatic tools allow for the amplification and subsequent accurate 
characterization of the sequence of monomers. Metabolomics, in contrast, aims 
to identify and elucidate the structure of metabolites, which are not sequences 
of monomers and do not result from a residue-by-residue transfer of information.
Instead, the large diversity of metabolites in living organisms results from 
series of chemical transformations catalyzed mainly by enzymes.

As for the identification of proteins in proteomics, structural annotation
of metabolites in complex biological mixtures relies on tandem mass spectrometry
(MS/MS) analysis. However predicting MS/MS spectra for metabolites is much more 
challenging than for peptides. In practice, therefore, annotating metabolites
relies on their MS/MS spectra being present in reference databases 
\cite{Patti12,Rojas-Cherto12,Nikolskiy13,Tautenhahn12}.
In the simplest situation, the sample metabolite and its MS/MS spectra are
already included in a reference database, so that the metabolite is annotated
by matching both the intensities and the mass-to-charge values of each fragment 
ion to values from pure standard metabolites in the database.

Unfortunately, MS/MS spectra of a large number of known metabolites are not
described in databases \cite{Vinaixa16}. To assist the structural annotation of such metabolites,
efforts have emerged recently to heuristically predict fragmentation patterns
\textit{in silico} and compare these to experimental MS/MS spectra 
\cite{Duhrkop15,Heinonen12,Menikarachchi12,Wolf10,Gerlich13,Li13,Ridder14,Allen14}.
Other methods are not based on MS/MS data and instead use the accurate mass of MS
peaks. These methods often require additional information about the sample metabolite
(e.g. pathways in which they participate), and the use of high precision
instruments and techniques such as the Fourier Transform Ion Cyclotron
Resonance (FTICR) or Orbitrap Fourier transform MS 
\cite{Breitling06,Rogers09,Weber10}.
Despite 
these efforts, the false positive rate of these methods is still too high to use
them in untargeted metabolomics analysis.

Finally, in the most challenging case (which, arguably, is also very common),
the sample metabolite is completely unknown, that is, the metabolite is not
described in databases \cite{Kalisiak09}.
Existing approaches for this situation use neutral 
losses and characteristic fragment ions as signatures for unique chemical
functional groups. These approaches have proved to be effective for classifying 
very specific lipid structures such as acyl-carnitines (e.g., fragments at m/z
85,02 and 60,08), glycerolipids, glycerophospholipids (e.g., fragment 
m/z 184.07) and sphingolipids 
\cite{Kind13,Ma14,Lynn15}.
However, there is not a general tool that
allows structural annotation of unknown metabolites from their MS/MS spectra. 

\begin{figure}
\centerline{\includegraphics[width=0.9\textwidth]{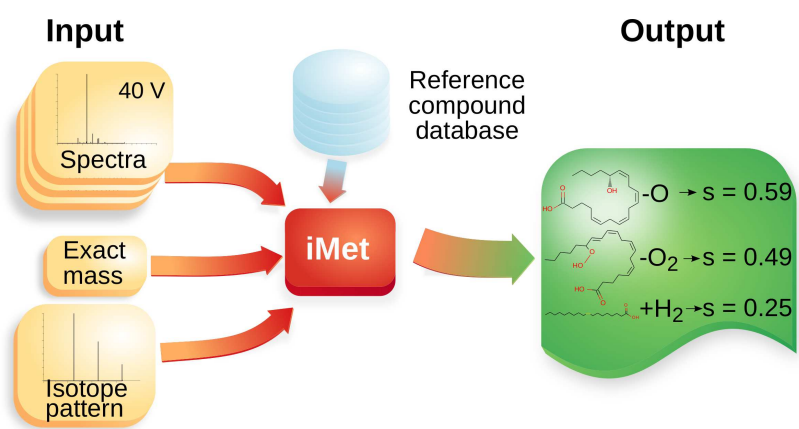}}
\caption{The iMet algorithm.
For each unknown metabolite, the algorithm 
takes as input its experimental MS/MS spectra (for different collision 
energies), its exact mass, and (optionally) its isotopic pattern. Then, iMet compares 
these inputs to a reference database containing experimental MS/MS spectra 
of known compounds (see text and Fig. \ref{fig:05} for details).
The algorithm outputs 
a sorted list of candidate neighbors of the unknown metabolite. For each 
candidate neighbor, iMet gives a chemical transformation that converts the 
candidate into the unknown metabolite. The accuracy of the prediction is 
defined as a numeric score ($s$), whose value goes from 0 for the lowest 
reliability to 1 for the highest reliability.}
\label{fig:01}
\end{figure}

To help in the annotation of completely unknown metabolites
we have developed iMet, a computational tool designed to fill that gap
(Fig.~\ref{fig:01}; iMet is avilable online at http://imet.seeslab.net).
Its two only inputs are the ESI Q-TOF MS/MS spectra and the exact mass of an
unknown metabolite (i.e., a metabolite that is not yet annotated in 
any database). Optionally, to increase the accuracy, the isotopic
pattern of the intact unknown ion can also be supplied.
Given these
inputs, the algorithm identifies metabolites in a reference database that are likely to be structurally very similar to the unknown metabolite. Finally, iMet produces a sorted
list of candidates, ranked by their similarity to the
unknown metabolite. The algorithm also suggests the chemical transformation that is most likely to separate each of the candidates from the unknown metabolite.
% figure 1

\section{Basic principle of iMet}

Metabolites can be represented as nodes in a network; two metabolites A and B
are connected, that is, are neighbors, if one can obtain the chemical structure of B by 
a chemical transformation of A, and vice versa (see Fig \ref{fig:02}A).
By a
chemical 
transformation here we mean the addition or removal of a moiety, or a 
conformational change. By definition, neighbor metabolites are structurally 
more similar than a typical pair of non-neighbor metabolites, and this 
structural similarity should be reflected in their MS/MS spectra because the 
fragmentation pattern of a metabolite highly depends on its chemical structure. 
Therefore, from the MS/MS spectrum of a metabolite that is not annotated in the 
network, a trained algorithm should be able to locate possible neighbors on the 
basis of spectral similarity.

To probe this idea, we built a network of neighbor metabolites on the basis
of 814 ``reactant pairs'' (RP) defined in the KEGG database 
\cite{Kanehisa04}. In KEGG, 
substrates and products of a known biochemical reaction are paired according to 
their chemical structure using graph theory \cite{Hattori03}
(see Fig \ref{fig:02}A).
By construction, 
the two metabolites in a RP are neighbors in the network, so we use RPs as 
ground truth for neighborhood. (Note, however, that not all neighbor metabolite
are annotated as RPs in KEGG. This occurs, for example, when there is no 
described biochemical reaction that can transform one into the other, even 
though they are structurally very similar. Thus, the network of RPs in KEGG 
and our dataset of neighbor metabolites is a subgraph of the full network of 
neighbor metabolites that potentially exist in nature.)

\subsection{Neighbor metabolites share structural similarities}

\begin{figure}
\centerline{\includegraphics[width=0.9\textwidth]{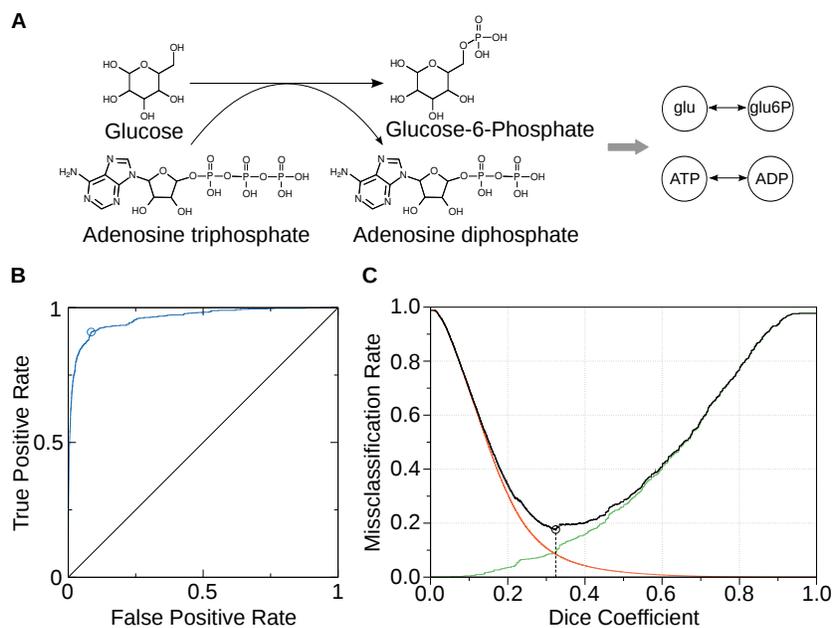}}
\caption{Neighbor metabolites.
(A) An example of neighbor metabolites. Phosphorylation of glucose. 
Glucose (glu) is transformed into glucose-6-phosphate (glu-6P), while 
adenosine triphosphate (ATP) is dephosphorylated into adenosine diphosphate 
(ADP). Following the definition of reactant pairs in KEGG, glu and glu-6P 
are one chemical transformation (phosphorylation) away from each other. 
The same applies to ATP and ADP. (B) ROC curve of the structural similarity
of neighbor and non-neighbor metabolites on the basis of the Dice coefficient.
The AUC is of 0.96, indicating that
neighbor metabolites have higher structural similarity than non-neighbor
metabolites. The blue circle marks the maximum discrimination point.
(C) False Negative Rate, i.e. ratio of RPs with a Dice coefficient below
(green) a certain Dice coefficient value,
and False Positive Rate, i.e. ratio of non RPs with
a Dice coefficient above (red) a certain Dice coefficient value, as a
function of that value. Black line corresponds to the sum of the other
two curves. The black circle indicates the minimum of this curve,
which corresponds to the maximum discrimination point. }
\label{fig:02}
\end{figure}

To assess the structural similarity between neighbor metabolites,
we computed the Dice coefficient \cite{Dice45}
between ECFP4 molecular fingerprints
\cite{Rogers2010} of pairs of metabolites.
We obtained the molecular fingerprint for 5,060 different metabolites
from public databases,
including 932 metabolites listed in KEGG as forming a RP with another
metabolite (and thus, being effectively a neighbor of another metabolite
in the network). We compared all the possible pairs of structures and computed the
receiver operating characteristic (ROC) curve \cite{Hanley82},
to check if structural similarity could be used to discriminate between neighbors
and non-neighbor metabolites
(Fig \ref{fig:02}B).
An overall measure of the discriminatory power is the area under 
the ROC curve (AUC statistic), which indicates how often RP (and in general,
neighbor metabolites) have a 
higher structural similarity than metabolites that are not RPs. The value of
the AUC found in this case was of 0.96, indicating that in the vast majority
of cases, two metabolites that are neighbors (that are one 
chemical tranformation away from each other) have a more similar structure
than those that are not. We also found that 95\% of the RPs have a Dice
coefficient higher than 0.22. 

With the aim of establishing the best threshold in a scale of 0 to 1 to separate between RPs and non RPs, we looked for the Dice coefficient value that
minimizes the classification errors. To do so,
we calculated the False Positive Rate (FPR) and the False Negative Rate (FNR)
at each
value of the Dice coefficient. In this context, the FPR
corresponds to the proportion of non RPs that have a higher Dice coefficient value
than a given threshold, while the FNR represents the
ratio of RPs that have a Dice coefficient value lower than that same
threshold (Fig \ref{fig:02}C). Adding these two curves into a third
curve, we obtain the missclassification ratio, that is the
ratio of pairs of metabolites that
would be incorrectly classified if we used a certain value of their Dice
coefficient to discriminate between RPs and non RPs. 
We found that the missclassification ratio is minimal
for a Dice coefficient value of 0.32, with a FPR of 0.09 and a FNR of 0.09, for
a total missclassification ratio of 0.18 (highlighted in 
Fig \ref{fig:02}B and C).
By using this value as a threshold to separate between RPs and
non RPs the classification error is minimum.

%figure 2

\subsection{Neighbor metabolites have similar MS/MS spectra}
In order to numerically quantify the similarities between two MS/MS spectra, 
we use the cosine similarity, as this method is both
time efficient and accurate (see Supplementary Data and Supplementary Table S1
for a discussion).
In particular, we discretize each spectrum in 
equal intervals of width $\delta m$. In this way, for each spectrum we can 
construct an intensity vector $\mathbf{v}$ in which element $v_i$ corresponds
to the relative intensity of m/z values in the interval [$m_i, m_i + \delta m$]. (Note that we use $\delta m$ = 0.01Da, and we disregard relative intensity values
below 1\% of the highest m/z value.)
Then, the cosine similarity $c$ between
spectra $\mathbf{v}$ and $\mathbf{u}$ is simply the dot product of the two vectors divided by the
product of their norms.

\begin{equation}\label{eq:c}
c = \frac{\sum\limits_{i} v_i u_i}{\|\mathbf{v}\| \|\mathbf{u}\|}
\end{equation}

\begin{figure}
\centerline{\includegraphics[width=.9\textwidth]{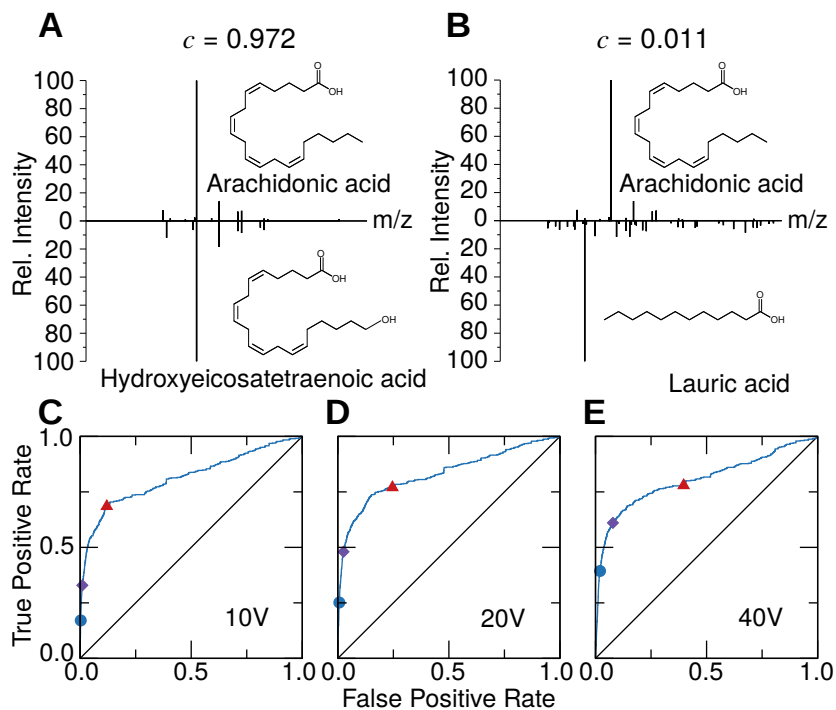}}
\caption{Similarity of MS/MS spectra discriminates between neighbor and 
non-neighbor metabolites.
(A) MS/MS spectrum similarity for two neighbor 
metabolites (spectral similarity 0.972), and (B) for two non-neighbor 
metabolites (spectral similarity 0.011). (C-E) Classification power of 
the cosine similarity.
We show the ROC curve for 
the cosine similarity when discriminating between RPs and non RPs in KEGG for
different collision energies (10, 20 and 40V) 
in negative ionization mode, with a total area under the curve of (C) 0.81, 
(D) 0.83, and (E) 0.80, respectively. The three highlighted symbols correspond to cosine 
similarities of 0.5 (blue dot), 0.1 (purple diamond), and the first non-zero value (red triangle).}
\label{fig:03}

\end{figure}

To validate the hypothesis that spectral similarity is indicative of neighborhood in the network, we quantified to which
extent metabolites that are neighbors have similar MS/MS spectra
(Fig.
\ref{fig:03}).
To this end, 
we considered all metabolites in KEGG for which we had the experimental MS/MS 
spectra from public databases (HMDB \cite{Wishart13}, MassBank \cite{Horai10}
and METLIN \cite{Smith05}), 
which corresponds to 1,763 metabolites, and compared their spectra using the 
cosine similarity (Fig.
\ref{fig:03}A and B). 

We used the ROC curve to quantify the power of spectral 
similarity to distinguish pairs of 
metabolites that are RPs in KEGG (and, therefore, neighbors) from those that are
not (Fig. 
\ref{fig:03}C, D and E).
In this case, the area under the ROC curve indicates how often RP metabolites have a 
higher spectral similarity than metabolites that are not RPs. For the three collision 
energies 10V, 20V and 40V in negative ionization mode, we found 
AUC$_{\text{10V}}$ = 0.81, AUC$_{\text{20V}}$ = 0.83 and 
AUC$_{\text{40V}}$ = 0.80, which indicates that the similarity between MS/MS 
spectra is useful to identify 
neighbor metabolites. Comparing MS/MS spectra in 
positive ionization mode gave similar AUC values. 
% figure 3
Note that these metrics quantify the predictive power of the spectral
similarity when comparing only two MS/MS spectra with the same collision
energy and ionization mode.

In general, MS/MS spectra acquired at high collision energies have higher 
sensitivity but lower specificity, and conversely for low collision energies. 
This implies that comparing two spectra obtained at high collision energies 
results in a highly conservative classification method, discarding pairs that 
are actually neighbor metabolites (low specificity or true negative rate), but 
assuring that most of the metabolites classified as neighbor metabolites are 
real neighbors (high sensitivity or true positive rate). In contrast, spectral
similarity becomes a poorer classification method at low collision energies: 
while most of the neighbor metabolites are correctly classified as such, some 
non-neighbor metabolites are also labeled as neighbors. Finally, 
the anal\-y\-sis 
reveals that information is usually non-redundant--some pairs of 
metabolites have high spectral similarity at high collision energies and low
similarity at low energies, whereas for other pairs the opposite is true.

These results indicate that, indeed, spectral similarity at a fixed collision energy is predictive to some extent. As we show next, however, the predictive power of spectral similarity can be increased by considering spectra at different collision energies simultaneously, and combining them with mass difference, and, optionally, the isotopic pattern of the
unknown metabolite (i.e., precursor ion).

\subsection{Neighbor metabolites have well-defined mass differences, which 
correspond to well-defined chemical transformations}

\begin{figure}
\centerline{\includegraphics[width=.9\textwidth]{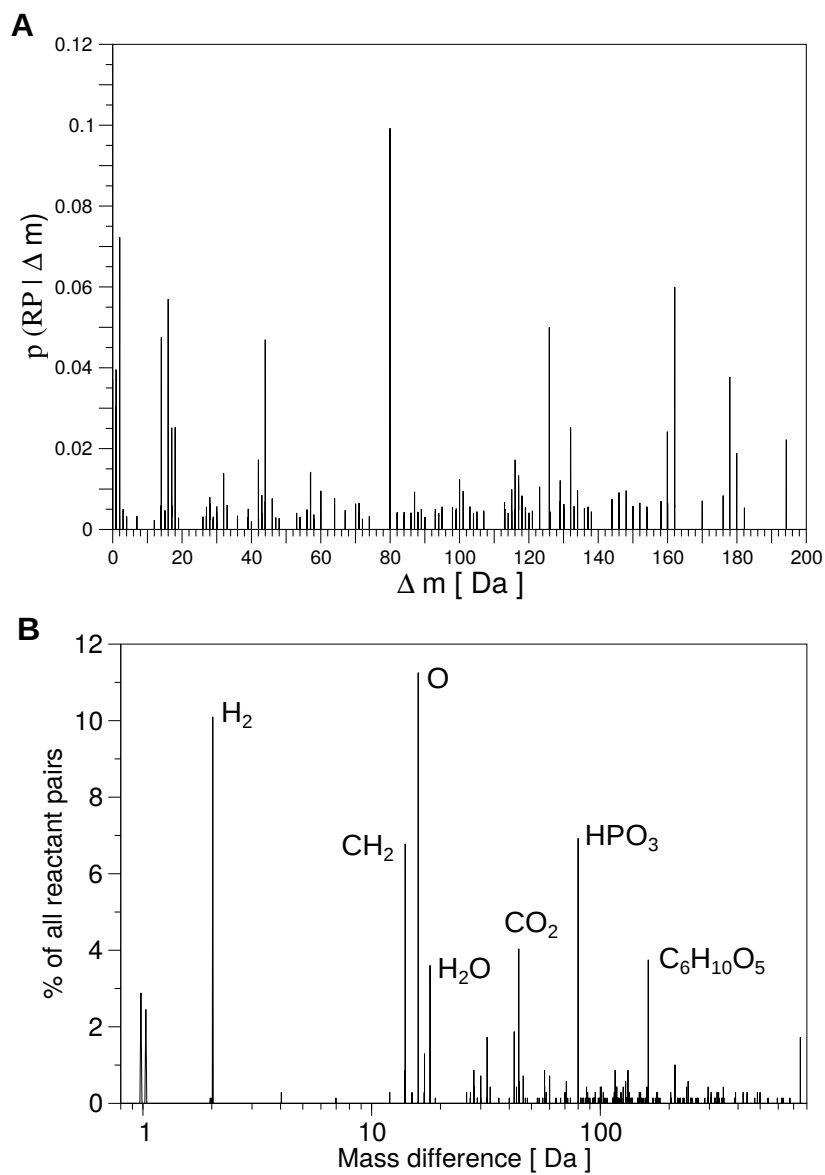}}
\caption{Probability of neighborhood and of chemical transformation.
(A) Probability of RP given a mass difference between two metabolites. 
We show the fraction of RPs that have associated a mass difference within 
a specific interval. We constructed the histogram using all compounds listed 
in KEGG, with bins of 0.01 Da. (B) Percentage of all 
reactant pairs in 
KEGG with a certain mass difference. The most frequent mass differences 
correspond to well defined moieties. The seven most frequent moieties 
(highlighted in the Figure) account for 46\% of all reactant pairs.}
\label{fig:04}
\end{figure}

To complement the information obtained from the spectral similarity, we 
study the differences in exact mass between neighbor metabolites 
(Fig. 
\ref{fig:04}).
The mass difference between two metabolites corresponds to the mass of the group 
of atoms added to (or removed from) one of the metabolites to convert it into the other. 

As before, we take the mass difference, $\Delta m$, between KEGG RPs as ground 
truth.
We considered 5,060 different metabolites including 814 RPs.  
Even though the KEGG database comprises more RPs,
the limiting factor is how many of these RPs have MS/MS
spectra in public databases. We are convinced that adding more metabolites
(165 by December 2015) to the analysis (out of a total of 5,060 metabolites
currently considered) would not alter the results presented other
than increasing its accuracy.
Calculating the mass difference for every pair of metabolites in our database
and plotting the proportion of RPs for each value of the mass difference 
yields a histogram that we assume reflects the probability of two metabolites being 
neighbors given their mass difference, 
although specific systems may deviate from this general (average) pattern
\cite{Morreel14}.

As we show in Fig.
\ref{fig:04}A, 
this 
distribution displays well-defined maxima at specific $\Delta m$ values. 
Therefore, it is much more likely that two metabolites are neighbors if their 
$\Delta m$ corresponds to one of the maxima of the distribution in Fig.
\ref{fig:04}A.
% figure 4
To understand what these $\Delta m$ represent, we extracted 202 distinct 
chemical transformations from a total of 814 reactant pairs,
(see Supplementary Table S2). 
Since each chemical transformation implies 
a well-defined 
mass difference, the distribution of mass differences among RPs is localized 
around certain values that correspond to the most common interconversions of 
atoms (Fig.
\ref{fig:04}B).
For example, 11.3\% of all reactant pairs correspond
to the net addition of an oxygen atom ($\Delta m$=15.995 Da), 10.1\% to the net
addition of H$_{\text{2}}$ ($\Delta m$=2.016 Da), and 6.9\% to the addition 
of a phosphate group ($\Delta m$=79.966 Da). In summary, a relatively small 
number of transformations account for a large number of the observed RPs (for example, the seven most common 
transformations account for 46\% of the pairs).

\subsection{A random forest classifier identifies neighbor metabolites on 
the basis of mass difference and MS/MS spectral similarity}

Taken together, our results indicate that MS/MS spectral 
similarity and mass difference are both predictive of neighborhood between two metabolites. 
Therefore, given an unknown metabolite and a library with MS/MS spectra,  
we propose that it is possible to identify the metabolites 
in the library that are most likely to be neighbors of the unknown one.
 
To demonstrate this, we implemented a random forest classifier \cite{Breiman01}
to identify 
potential neighbors. The random forest classifier has the advantage of 
automatically taking care of the non-monotonic relationship between mass 
difference and probability of neighborhood, as well as the complex non-linear 
similarity patterns between MS/MS spectra at different collision energies. We 
trained the classifier using 50,000 metabolite pairs, including all RPs 
for which we have MS/MS spectra and completed with randomly chosen pairs among a library of 5,060 compounds with MS/MS data in databases such as 
HMDB \cite{Wishart13}, MassBank \cite{Horai10}, and METLIN 
\cite{Smith05}. Increasing the size of the training set slowed the training
to the point of making it unfeasible, and did not result in an increase of the
accuracy of the classifier (because adding more pairs of unrelated compounds
did not provide new useful information to the model).

Based on the accurate mass-to-charge (m/z) measurement of a protonated 
(M+H)$^{\text{+}}$ or deprotonated (M-H)$^{\text{-}}$ precursor ion of the 
unknown metabolite (mass error \textless 0.005 Da), its MS/MS spectra, and 
its experimental isotopic distribution when available, the trained classifier 
yields a sorted list of candidate neighbors of the unknown metabolite 
(Fig. \ref{fig:01}), 
chosen from amongst the 5,060 compounds included in our database. Moreover, 
iMet uses the most common chemical transformation between RPs 
(Fig.
\ref{fig:04}B)
to 
predict the unknown metabolite's chemical formula.

All in all, iMet outputs a sorted list of candidate neighbors of the unknown 
metabolite on the basis of mass difference and MS/MS spectral similarity. For 
every candidate, and given its mass difference with the unknown metabolite, 
iMet gives the chemical transformation (group of atoms) that converts the 
candidate into the unknown metabolite. The reliability of the prediction is 
given as a numeric score ($s$), whose value goes from 0 for the least reliable 
to 1 for the most reliable.

\section{Materials and Methods}
\subsection{Construction of the classifier}
We used the R package ``randomForest'' for the classifier \cite{Liaw02}.
We trained the
classifier with a dataset that contains all the RPs in the KEGG database for
which we have MS/MS spectra (814 pairs), plus an additional 49,186 randomly
chosen pairs of metabolites. Therefore, we trained the algorithm with 50,000
pairs of metabolites, a set deliberately enriched with RPs. 
We need the training compounds to be in KEGG because we use KEGG reactant
pairs as our ground truth for compound neighborhood.
Increasing the size
of the training set did not alter significantly the results obtained. The
classifier uses the following features:  (i) the cosine similarity between the
MS/MS spectra of the two metabolites at all available collision energies; (ii)
the fraction of metabolite pairs with the observed mass difference that are
actually reactant pairs according to the KEGG database. The classifier tries to
predict whether the pair of metabolites are neighbors or not.

Our classification algorithm takes also into account experimental mass errors.
Specifically, we introduced a shift in the exact mass of every metabolite of the
training dataset, changing its mass to a value randomly drawn from a Gaussian
distribution centered around the exact mass of the metabolite and a standard
deviation of 0.0025Da. This way, the algorithm correctly deals with the
experimental error of the unknown target metabolite.

\subsection{MS/MS database}
Our database is composed of 29,242 MS/MS spectra from 5,060 different compounds
obtained from the databases HMDB \cite{Wishart13},
MassBank \cite{Horai10} and METLIN \cite{Smith05},
obtained with a Q-TOF instrument and at different collision energies and
ionization modes.
Although it is true that HMDB contains ~45.000 compound entries, only ~8\%
of those compounds have electrospray ionization MS/MS spectra. A similar
percentage of the compounds in METLIN have this type of spectra. All in all,
with electrospray ionization MS/MS spectra for 5,060
compounds, our reference database is about as comprehensive as one can 
get nowadays \cite{Vinaixa16}.

\subsection{iMet step by step}
\begin{figure}
\centerline{\includegraphics[width=0.5\textwidth]{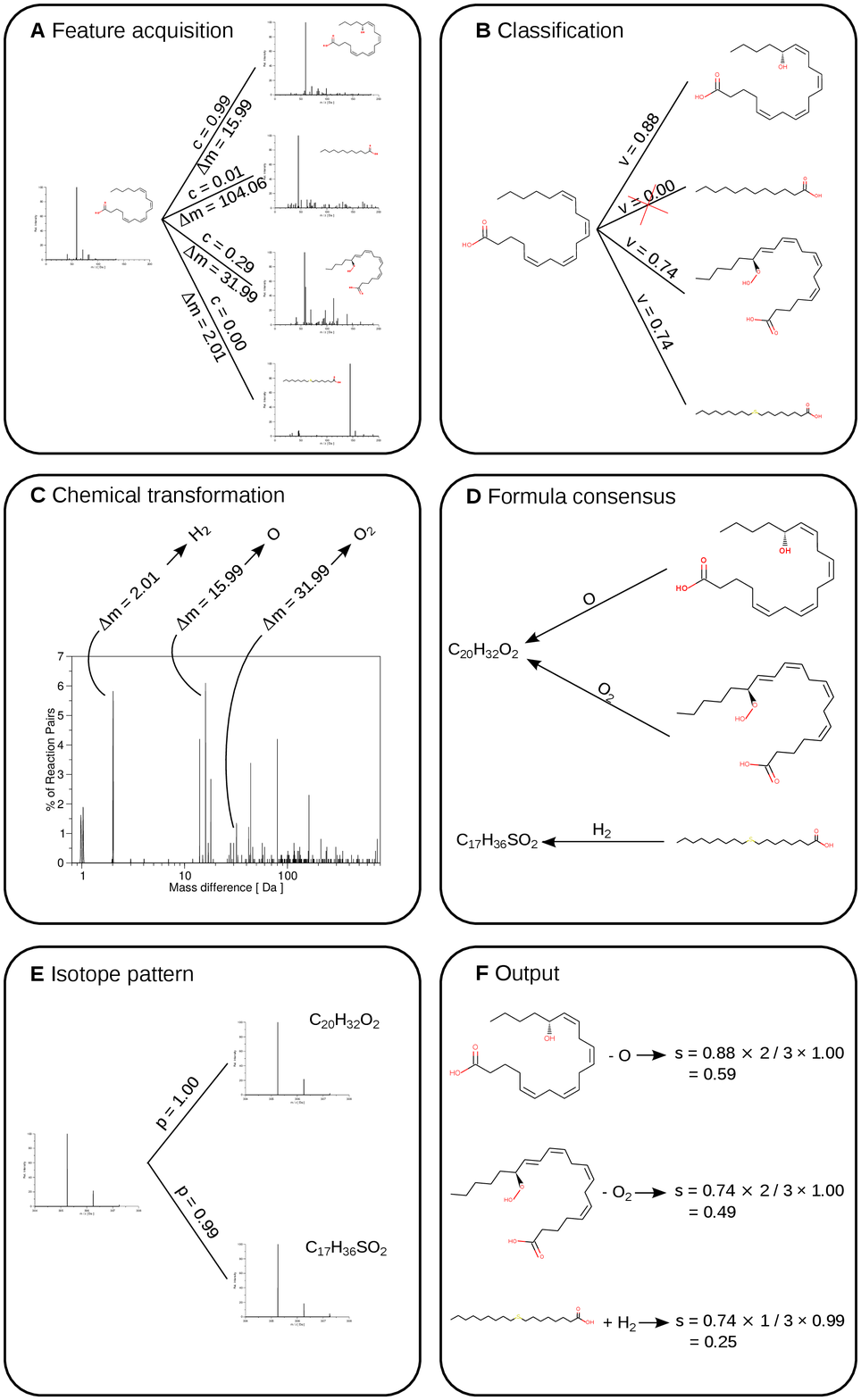}}
\caption{Schematics of the general procedure followed by the iMet algorithm. 
In this simplified example the ``unknown'' test metabolite is arachidonic acid, 
and the database consists only of four metabolites (16-hydroxyeicosatetraenoic 
acid, dodecanoic acid, 15-hydroperoxyeicosatetraenoic acid and 9-thiastearic 
acid) with their corresponding MS/MS spectra obtained from the same 
experimental conditions. All mass values are expressed in Da.
(A) Feature acquisition. The MS/MS 
spectrum of the test metabolite is compared to those in the database, obtaining 
a spectral similarity ($c$) and a value of mass difference.
(B) Classification. Using the above-mentioned features, the classifier
computes the likelihood of every compound in the database of being a neighbor 
of the sample metabolite ($v$). Those metabolites with a likelihood higher than 
0.5 are candidate neighbors of the test metabolite. 
(C) Determination of chemical transformation. For each candidate, a moiety is 
proposed according to the mass difference between the candidate neighbor and 
the test metabolite. 
(D) Formula consensus. The molecular formula of the test 
metabolite is then calculated as the sum of the candidate neighbor's formula 
plus the chemical transformation. The frequency of each formula amongst the 
formulas of all the candidate neighbors is computed ($f$). 
(E) Isotope pattern. 
If the experimental isotope pattern of the test metabolite is 
available, the algorithm compares it to the theoretical isotope pattern of the 
proposed formula for the sample metabolite ($p$). 
(F) Output. The final output 
is a sorted list of candidates, ranked by the final algorithm score ($s$). 
}\label{fig:05}
\end{figure}

Given an unknown metabolite $\chi$, a reference set of $Q$ metabolites
$R= \{R_1, R_2 \ldots, R_Q\}$, and a trained random forest (see ``Construction
of the classifier'' above), the algorithm builds a list of candidate neighbors
of $\chi$ as follows (Fig. \ref{fig:05}):
\begin{enumerate}
\item{Obtaining spectral similarities and mass differences:
Given an unknown metabolite $\chi$, the algorithm computes the cosine similarity
between the unknown metabolite's ESI Q-TOF MS/MS spectra and every MS/MS spectra in
the database obtained under the same experimental conditions (i.e., ionization
mode and collision energy). 
Spectral similarity is calculated for each collision energy separately,
that is, we do not combine or merge spectra from different collisions energies.
Our algorithm can work using just one or two collision energies, but it
performs better if all possible collision energies are provided as input data.
The algorithm also computes the exact mass of
$\chi$, as the precursor ion mass plus or minus the hydrogen atom mass,
depending on the ionization mode ([M+H]$^+$ or [M-H]$^-$, for positive or negative
ionization, respectively). Although other possible adducts can be formed,
the protonated adducts are by far the most abundant
in ESI-MS and their MS/MS spectra are prevalent in databases \cite{Vinaixa16},
and so iMet only
takes into account protonated adducts (although application to other adducts would be immediate).

With the exact mass M$_S$ of the unknown metabolite, the algorithm computes
the mass difference $\Delta m$ between $\chi$ and every metabolite in the
reference set $R$. In each case, the algorithm obtains the fraction of all
metabolite pairs with that $\Delta m$ that are actually reactant pairs.
We assume that this ratio reflects the probability of two metabolites being 
neighbors
given only their mass difference (Fig.
\ref{fig:04}A).}

\item{Classification: Using the spectral similarity obtained from the
comparison of MS/MS spectra and the ratio of RPs from the calculated mass
difference, the classification algorithm computes the likelihood for each
metabolite in the database of being neighbors with the unknown metabolite
(score $v$). Those metabolites in the database with a likelihood lower than
0.5 are discarded; the remaining metabolites are the most likely to form a
RP with the unknown metabolite. These are the candidate neighbors of the
unknown sample.}

\item{Determination of the chemical transformation: From the mass difference
of each candidate neighbor, the algorithm proposes the most probable chemical
transformations linking each candidate to the unknown metabolite. Specifically,
the algorithm considers all possible molecular formulas that are compatible
with the observed mass difference. Note that, in general, these molecular
formulas coincide with one of the 202 unique chemical transformations derived
from RPs listed in Supplementary Table S2.}\label{itm:biotrans}

\item{Formula consensus: Given that the molecular formulas of the candidate
neighbor and its associated chemical transformation are known (from step
(\ref{itm:biotrans})), the algorithm postulates a final formula for the
unknown metabolite based on the sum of these two formulas. The algorithm
then computes the frequency of each potential final formula as the proportion
of candidate neighbors and their chemical transformation that yield that same
molecular formula. Then, in order to resolve possible ties, the algorithm
favors simple chemical transformations over complex ones by dividing the
frequency of each final formula by the square root of the number of atoms
associated with the chemical transformation. This final number is the formula
consensus score (score $f$).}
\item{Isotope pattern comparison: The theoretical MS isotope pattern of every
final formula is computed (as described in ``Elucidation of the theoretical
isotope pattern'' in the Supplementary Data),
and compared to the experimental isotope pattern of the
unknown metabolite. As before, we use the spectral similarity previously
defined. Consequently, the algorithm associates every candidate neighbor
with a spectral similarity (score $p$) based on its isotopic distribution.}

\item{Output: for each unknown metabolite-candidate neighbor pair, we have
an overall score ($s$)

\begin{equation}\label{eq:score}
s=v \times f \times p
\end{equation}
where, again, $v$ is the score of the RF classifier, $f$ the formula
consensus score, and $p$ is the spectral similarity between theoretical and
empirical isotopic patterns.}
\end{enumerate}
All in all, the outcome of our algorithm is a sorted list of candidate
neighbors of the unknown sample $\chi$, ranked by their score $s$.
% figure 5

\section{Results and discussion}

\subsection{Cross-validation of iMet using 148 test metabolites}

\begin{figure}
\centerline{\includegraphics[width=.9\textwidth]{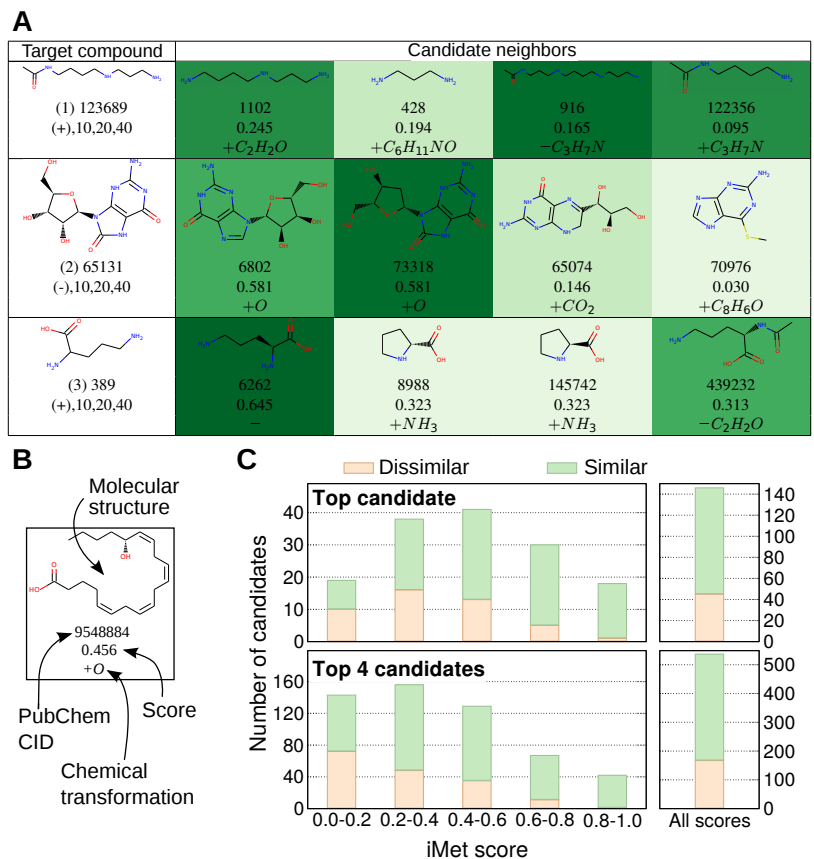}}
\caption{Cross-validation of iMet.
(A) Results for three naturally 
occurring metabolites. The test metabolites, which are unknown to the 
algorithm, are shown in the first column along with their PubChem compound 
ID \cite{Bolton08} (123689, N(8)-acetylspermidine;
65131, 8-hydroxyguanosine; 
389, ornithine), the sign of the ionization as well as the collision 
energies used to obtain the MS/MS spectra. The other columns contain the 
top four candidate neighbors ranked from highest to lowest score, along 
with their PubChem compound ID, the iMet score, and the proposed chemical 
transformation.
Each candidate neighbor metabolite is colored according to 
the value of the Dice coefficient calculated between its own molecular fingerprint
and that of the target metabolite, ranging from white (Dice value of 0)
to dark green (for a Dice value of 1).
(B) Legend for the candidate 
neighbors' columns in (A). (C) Summary of validation for 152 tests 
(see Supplementary Figures S1 and S2).
We display the number of 
top candidate neighbors that are structurally similar to the test metabolite 
as a function of the iMet score. Only when the iMet score is very low 
(0.0-0.2), a significant fraction of candidate neighbors are not structurally 
similar to the test metabolite; for larger iMet scores, candidate neighbors are 
overwhelmingly good structural matches to the test.}
\label{fig:06}
\end{figure}

To validate iMet, we experimentally obtained in our laboratory 
MS/MS spectra of 48 metabolites 
in different conditions, for a total of 52 different tests
as some metabolites are tested in both positive and negative
ionization modes separately
(all the test spectra can be found in Supplementary File 1;
see also Supplementary Table S3 for cross-references in different
databases of the test metabolites used).
To ensure structural and biochemical diversity of tests,
these include nucleotides and nucleosides, both natural and unnatural amino 
acids, 
vitamins, sphingolipids, polyamines and fatty acids, among others
(see Supplementary Table S4 for a complete listing of pathways
covered by these tests).
We excluded their spectra from the training set and manually removed their entries 
from our database, effectively turning them into unknown compounds for the 
purpose of validation.
For these 48 metabolites, we ran iMet
against a reference database of 5,012 compounds (see ``MS/MS database'' above)
and evaluated the quality of each prediction. We considered two metabolites
(the unknown metabolite and the candidate output by iMet) to be structurally 
similar if the Dice coefficient of their molecular fingerprints was above 0.32.
In these validation tests, 78\% of the top candidates identified by iMet were,
indeed, structurally similar to the target,
with the correct chemical transformations (Fig. \ref{fig:06}, see also
Supplementary Table S5 for complete results). 
For 91\% of the cases, at least one of the top four candidates 
suggested by iMet was structurally similar to the target, and the proposed 
chemical transformation was also correct.
% figure 6

We further validated the algorithm using 100 randomly selected metabolites 
(see Supplementary Table S3 for cross-references and 
S6 for the pathway coverage of these tests)
whose MS/MS spectra we took directly from our 
reference 
MS/MS database of 5,060 metabolites. We followed a leave-one-out 
cross-validation procedure, so each metabolite was tested individually removing 
it from the database. In this validation, 65\% of the top candidates were 
structurally similar (Dice coefficient > 0.32) to the unknown test metabolite,
67\% of
candidate neighbors among the top four are structurally similar to the test
metabolite, and in 88\% of the cases 
the algorithm predicted that one of the top four candidates was structurally 
similar to the test metabolite (see Fig. \ref{fig:06} and Supplementary Table 7
for the complete table of results). 

Overall, combining both cross-validation experiments, iMet was able to correctly
identify the unknown test metabolite from the top candidate in 69\% of the 
cases. 
In 89\% of the cases at least one of the top four candidates was structurally 
similar to the test metabolite. In 88\% of the cases, the top formula proposed 
by iMet was the correct formula of the test metabolite.

\subsection{The CASMI challenge}

To test iMet under the most challenging circumstances, we tested it using
those metabolites proposed in the Critical Assessment of Small Molecule 
Identification (CASMI) challenges from years 2012-2014.
We downloaded the spectra for 32 metabolites obtained using an ESI-QTOF
mass spectrometer. From these, 26 metabolites
were obtained using other collision energies than those
used in the training set (10, 20 or 40V). We tested them nevertheless
to evaluate the performance of iMet when confronted with
spectra obtained using inaccurate experimental 
data (for example, we introduced spectra obtained at 25V as if they were 
obtained at 20V; or 35V spectra as if they were 40V).
For these 26 metabolites, 60\% of
the top candidates suggested by iMet were structurally similar to the
test metabolite, and in 70\% of the results iMet was able to locate at
least one structurally similar metabolite in the database
(see Supplementary Table S8). 

When including in the test metabolites from the CASMI challenges that 
were obtained using collision energies of 10, 20 or 40V
(for a total of 45 different tests),
iMet located at least one structurally similar
metabolite in 68\% of the tests.
59\% of the top candidates were
structurally similar (Dice coefficient > 0.32)
to the unknown test metabolite. 
These results suggest that iMet does not decrease substantially 
its accuracy when using 
slightly ``erroneous'' collision energies as inputs.

\subsection{Conclusions}

We have demonstrated that experimental MS/MS spectra can be used to their full advantage
for the structural annotation of unknown (i.e., undiscovered) metabolites, 
unlike previous approaches that use tandem MS spectral similarity 
\cite{Rojas-Cherto12,Nikolskiy13,Tautenhahn12}
or \textit{in silico} predictions of MS/MS spectra to identify known metabolites 
\cite{Heinonen12,Menikarachchi12,Wolf10,Gerlich13,Li13,Ridder14,Allen14,Kind13}.
In terms of MS/MS information, we have systematically demonstrated that 
MS/MS spectral similarity has enough discriminatory power
to distinguish between neighbor metabolites from non-neighbor metabolites.

Our cross-validation demonstrates that iMet is only limited by the experimental 
MS/MS library against which the unknown metabolite is compared, and to a lesser 
extent the coverage of reactant pairs and the space of chemical 
structure transformations. 
In particular, iMet will fail to annotate correctly an unknown metabolite
when no structurally similar metabolites are present in the reference
database. With databases having MS/MS spectra for only 8-10\% of their
compounds \cite{Vinaixa16},
this may seem a serious limitation. However, a simple calculation
suggests otherwise. Indeed, if each metabolite has, say, 10 neighbors, 6 of
which are known, then there are roughly 4 times as many unknown metabolites
(neighbors of known metabolites) as known metabolites; and there are roughly
4x10 times as many metabolites that are 2 biotransformations away from the
known ones. One can therefore see that an exponentially large number of new
metabolites can be reached from even a comparatively small set of known ones,
which is a well understood fact that has been well studied within the network
science literature (and, in particular, within the literature on the
structural properties of metabolic networks).
We anticipate that as the number of MS/MS spectra from known 
metabolites rises in public databases, so will do the predicting power of iMet.
 
Regarding chemical transformations, our network of reactant pairs is restricted 
to those biochemical reactions described in the KEGG database, which does not 
account for all chemical transformations occurring at any biological system. The
KEGG is, however, to our knowledge the only database that systematically shows 
paired substrates and products according to their structure transformations 
using graph theory \cite{Hattori03}. The significance of using this
information is that we 
can compute the probability of two metabolites being neighbors on the basis of 
the mass difference between them, without taking into consideration other 
attributes such as their chemical structures \cite{Hamdalla15}, functional 
groups, chemical 
reactivity or metabolic pathways \cite{Li13_2}. Rather, iMet uses a large set 
of chemical 
transformations described in biological systems to propose chemical formulas and
structures by adding (or removing) a group of atoms to a known chemical 
structure. This concept of chemical transformation is similar to that used 
previously by other computational approaches aiming to reduce the ambiguity in 
metabolite annotation
\cite{Breitling06,Rogers09,Weber10,Li13_2,Gipson08}.
Yet these previous 
approaches that do not make use of MS/MS data, require that the interrogated 
metabolite falls into a known metabolic pathway, and its chemical formula and 
structure must be known and described in a database 
\cite{Breitling06,Rogers09,Weber10,Gipson08}.

In summary, our algorithm has proven itself to be a unique tool in the 
annotation of unknown metabolites, as a stand-alone application.
iMet does not propose structures de novo, since mass spectrometry cannot
perform de novo identifications of chemical structures (unlike NMR).
iMet is intended to provide key information, such as the molecular formula
of the unknown compound, structurally similar (neighbor) compounds and the
moiety (if necessary) to transform the known neighbors into the unknown,
for organic chemists to propose candidate chemical structures based on
chemical knowledge \cite{Kalisiak09}.
We also acknowledge its potential 
when coupled to other applications such as MetFrag \cite{Wolf10},
MetFusion \cite{Gerlich13}, MAGMa \cite{Ridder14}, CFM-ID \cite{Allen14},
or MS2Analyzer \cite{Ma14}, for which the output of iMet could be used as 
inputs. This coupling would allow to circumvent the
need for a priori information of interrogated metabolites, as iMet would provide
a list of candidates without any other required information than its MS/MS 
spectra and its exact mass. By simulating the fragmentation pattern of those 
candidates, for example, the final result would be even more refined, achieving 
a higher accuracy in the identification of unknown metabolites.

\section*{Acknowledgments}

The authors would like to thank Mr. Manuel Miranda
for his help in the web implementation of the tool. 

\section*{Funding}

This work was supported by the
James S. McDonnell Foundation [220020228],
the Spanish Ministerio de Econom\'ia y Comptetitividad
[SAF2011-30578 and BFU2014-57466 to O.Y. and 
FIS2013-47532-C3 to A.A-M, M.S-P., R.G.],
and the European Union [PIRG-GA-2010-277166 
to R.G., PIRG-GA-2010-268342 to M.S.P. and FET-317532-MULTIPLEX 
to M.S.P. and R.G.].

\vspace*{12pt}
\noindent\textit{Conflict of interest:} None declared.
\bibliographystyle{unsrt}
\bibliography{iMet_no_format}
\end{document}